\documentclass[preprint,12pt,authoryear]{elsarticle}
\usepackage[english]{babel}
\usepackage[T1]{fontenc}
\usepackage{amsmath}
\usepackage{amsfonts}
\usepackage{amssymb}
\usepackage{graphicx}
\usepackage{epsfig}
\usepackage{float}
\usepackage{multirow}
\usepackage{url}
\usepackage[colorlinks=true, allcolors=blue]{hyperref}
\usepackage{ulem}
\usepackage{hyperref}
  
\usepackage{lineno}
\usepackage{rotating} 
\usepackage{lscape} 

\begin{document}

\title{Lifted particles from the fast spinning primary of the Near-Earth Asteroid (65803) Didymos}

\author{Nair Tr\'ogolo$^{1,2,3}$}
\author{Adriano Campo Bagatin$^{1,4}$ \footnote{acb@ua.es} } 


\author{Fernando Moreno$^5$}
\author{Paula G. Benavidez$^{1,4}$ }

\address{$^1$ Instituto de F\'isica Aplicada a las Ciencias y la Tecnolog\'ia, Universidad de Alicante, Spain. P.O. Box 99, 03080 Alicante, Spain}

\address{$^2$ Observatorio Astronómico, Universidad Nacional de Córdoba, Laprida 854, Córdoba X5000GBR, Argentina}

\address{$^3$ Consejo Nacional de Investigaciones Científicas y Técnicas (CONICET), Argentina}

\address{$^4$ Departamento de F\'isica, Ingenier\'ia de Sistemas y Teor\'ia de la Se\~nal, 
Universidad de Alicante, Spain. P.O. Box 99, 03080 Alicante, Spain}

\address{$^5$ Instituto de Astrof\'isica de Andaluc\'ia - CSIC, Glorieta de la Astronom\'ia s/n, 18008 Granada, Spain}

\begin{abstract}


An increasing number of Near Earth Asteroids (NEAs) in the range of a few hundred meters to a few kilometres in size have relatively high spin rates, from less than 4 h, down to $\sim$2.2 h, depending on spectral type. For some of these bodies, local acceleration near the equator may be directed outwards so that lift off of near-equatorial material is possible. In particular, this may be the case for asteroid Didymos, the primary of the (65803) Didymos binary system, which is the target of the DART (NASA) and Hera (ESA) space missions. The study of the dynamics of particles in such an environment has been carried out ---in the frame of the Hera mission and the EC-H2020 NEO-MAPP project--- according to the available shape model, known physical parameters and orbital information available before the DART impact. The presence of orbiting particles in the system is likely for most of the estimated  range of values for mass and volume.
The spatial mass density of ejected material is calculated for different particle sizes and at different heliocentric orbit epochs, revealing that large particles dominate the density distribution and that small particle  abundance depends on observation epoch. Estimates of take off and landing areas on Didymos are also reported. Available estimates of the system mass and primary extents, after  the DART  mission, confirm that the main conclusions of this study are valid in the context of current knowledge.

\end{abstract}

\maketitle
\section{Introduction}
Many Near Earth Asteroids (NEAs) have been discovered in the last two decades, including at sizes smaller than a few km and fast spin rates.  
\cite{Pravec2008} showed that an excess of slow rotators (spin period $T >$ 24 h), as well as fast rotators (2.2 h $<T<$ 2.8 h), is present in the NEA population at size $D >$ 0.2 km. The former is also found in main belt asteroids, while the excess of fast rotators is not, and it seems to be peculiar to NEAs. It is noteworthy that 2/3 of fast spinning NEAs are binaries \citep{Pravec2006} and they correspond to the concentration of fast spin rate of primaries of NEA binaries in front of the ``spin barrier'' at $\approx 2.2 $ h. 
The spin state with which asteroids enter the NEA region is mostly affected by non-catastrophic collisions while in the asteroid belt \citep{Holsapple2022}, and partially by the non-gravitational YORP effect. YORP is also the main driving source for NEAs spin up once in the inner planet region.

\subsection{Asteroid stability limits and spin--up}

We recall here the spin limits for mass shedding defining an upper threshold on the rotation rate at which a particle at the equator of a spherical body is at neutral equilibrium between gravity and centrifugal force. For a homogeneous sphere, the relation for the critical spin rate is:

\begin{equation}
\omega_{cr}=\sqrt{\frac{4 \pi G \rho}{3}}
\end{equation}

\noindent
where $\rho$ is the bulk density of the object. 
Spin limit depends on asteroid density: e.g., $T = 2.2$ h is the limit corresponding to a spherical body of  density $\rho=2250$ kg/m$^3$, which is typical of S-type gravitational aggregate asteroids; instead,   spin limit is  $T=3.0$ h for  C-type gravitational aggregate asteroids, with $\rho=1200$ kg/m$^3$. Such densities differ from the density of their meteorite analogues due to  macro--porous structure of gravitational aggregates.

As soon as enough spin data were available \citep{Pravec2000}, such a spin limit became evident when plotting size vs spin period. Very rarely asteroids larger than 200--300 m in diameter ($D$) have been observed with a rotational period smaller than 2.2 h. This has led to interpretations of the internal structure of such asteroids. Fast rotators are instead common at $D<200$ m.
Focusing the discussion on the size range 300 m --- 10 km, the interpretation of the spin limit leads to two slightly different regimes \citep{Holsapple2007}. For asteroids larger than 3 km, the spin barrier does not constrain whether these are strength--less objects or just cracked but coherent bodies. In fact, the upper limit on the tensile strength --given by the barrier itself-- is higher than a scaled tensile strength of cracked but coherent bodies. Instead, for asteroids smaller than 3 km, the maximum possible tensile strength allowed by the spin barrier is too low for these asteroids to be cracked but coherent bodies, so they should have predominantly cohesion--less structures. 

Non--coherent asteroids are aggregates that have re-accumulated fragments by self--gravity right after shattering  events. On the contrary, internally cracked objects may arise due to  shattering at the threshold energy for fragmentation, with little kinetic energy left to reshuffle fragments. Another way of producing cracked structures is by series of sub--catastrophic collisions summing up similar damage in the overall structure as one single barely shattering event \citep{Housen2009}. In that case, the object may be coherent allowing for some tensile stress.

Both coherent bodies and gravitational aggregates (GA) (often called rubble piles) may withstand spin rates higher than the critical ones for fluids found by Chandrasekhar \citep{Chandra1969}, and spin ideally up to the spin barrier around 2.2 h before failing apart. In the case of coherent --monolithic-- structure, that is due to internal solid state forces, which do withstand spin rates beyond that limit. This is especially the case of bodies smaller than 300 m.
Instead, non-coherent asteroids (as  gravitational aggregates are) may be spun up and undergo shape change corresponding to a minimum energy configuration --led by dissipative forces (e.g., internal friction)-- compatible with increasing angular momentum. That is achieved by rotation about the maximum angular inertia axis. As a result, some of those bodies might become top--shaped \citep{Cheng2021, Sabuwala2021}. 

Therefore, shear strength may be present due to friction and interlocking between GA components \citep{Richardson2002, Holsapple2007, Ferrari2020}, as a reaction to the shear stress due to centrifugal force, increasing structural yield. The presence of inter--particle cohesion has also been suggested \citep{Sanchez2012, Zhang2017, Zhang2021}, though still a matter of debate. Such shear strength, regardless of its nature,  may prevent the whole structure from falling apart when the rotation spin rate exceeds the stability limit for fluid bodies. This mechanism is successful until the spin barrier is reached. 
At that point, the body is no longer able to adjust the exceeding energy and angular momentum by shape change through energy dissipation by friction. Depending on internal stiffness, fission or mass shedding takes place, eventually leading to asteroid binary or pair/clan formation \citep{Pravec2019}. The mechanisms of formation of NEA binary systems are a matter of debate and are beyond the scope of this study.

 \subsection{NEA binary systems with fast spinning primary}

Many binary systems in the NEA population share a number of common features: (a) small mass ratio ($\approx $ 0.01) for satellite to primary components; (b) fast spin primary; (c) top-shape primary  \citep{Naidu2020, Roberts2021, Walsh2012}.
Of all binary systems with fast rotating primary, we identified a handful near the edge of stability, as reported in Table \ref{tab:tabNEAB} together with a number of single bodies. In this work, we focus on the Didymos system because of its interest as the target of  both DART (NASA) and Hera (ESA) space missions. However, the 1996 FG3 binary system also was the former goal of the MarcoPolo-R mission and is now the sample return target of the Chinese (CNSA)  Zheng He space mission.
Beyond binary primaries, it is worth mentioning that some lonely top--shape NEAs also show fast spin rates. This is indeed the case of 2008 EV5, which was the target of the un--selected ESA (2014) MarcoPolo-R mission. \\

\begin{table}[t]
\begin{center}
\begin{tabular}{ l  c  c c c}
\hline

Asteroid name & $D_p$ (km) & $D_s/D_p$ & $T$ (h) & Taxonomy type  \\  \hline
(65803) Didymos & 0.78 & 0.21 & 2.26 & S \\ 
(66063) 1998 RO1& 0.80 & 0.50 & 2.49 & - \\ 
(88710) 2001 SL9& 0.77 & 0.32 & 2.40 & Sr, Q \\
(164121) 2003 YT1& 1.10 & 0.19 & 2.34 & -\\  
(311066) 2004 DC& 0.36 & 0.19 & 2.57 & -\\ 
(137170) 1999 HF1 & 3.6 & 0.23 & 2.31 & X \\
(1862) Apollo   & 1.55 & 0.05 & 3.06 & Q \\
(175706) 1996 FG3 & 1.69 & 0.29 & 3.59 & C \\
(185851) 2000 DP107 & 0.80 & 0.38 & 2.77 & C \\
(276049) 2002 CE26 & 3.5 & 0.09 & 3.29 & C \\
          
\hline
\end{tabular}
\caption{Some NEA binaries with primaries  near the edge of stability. $D_p$ is the size of the primary, $D_s/D_p$ is the satellite/primary size ratio, and $T$ (h) is the primary spin period. (Source: Johnston's archive. https://www.johnstonsarchive.net/astro/asteroidmoons.html) }
\label{tab:tabNEAB}
\end{center}
\end{table}

 \subsubsection{The Didymos system}
 
The NEA binary (65803) Didymos is the S-type target of the DART (NASA) and Hera (ESA) space missions. Model predictions may be tested by combined data from those two missions, which makes this a particularly interesting system to study. This asteroid is classified as an Apollo NEA with a semimajor axis of 
$1.6442688843 \pm 1.6 \times 10^{-9}$ au, and a large eccentricity of $0.383882802 \pm 3\times 10^{-9}$ (pre-DART impact heliocentric ephemeris solution 181).
Its perihelion is therefore well inside the inner asteroid belt, where the asteroid spends 1/3 of its orbital period.

The pre-–impact estimations of the system main physical characteristics were the values available when this research started, and numerical simulations were run. 
The discussion on how post--DART impact estimation may affect results is carried out in Sec. 5. 

Didymos, the primary of the binary, had estimated principal axes extent sizes of $832\, (\pm 3\%)\times 837\, (\pm 3\%)\times 786\, (\pm 5\%)$ km, the last of which is the size along its spin axis. A Didymos shape model has been derived using both radar and optical telescope data, clearly indicating a top--shape \citep{Naidu2020} before the DART and LICIACube missions imaging. The mass of the system was estimated as $5.278 \times 10^{11} \pm 10\% $ kg \citep{Naidu2020} from the orbital period of the secondary, named Dimorphos, which is known, 11.9216289 $\pm$ 2.8 $\times 10^{-6}$ h (S. Naidu and S. Chesley, personal communication). 
The spin period of Didymos is $2.2600\pm 0.0001$ h \citep{Pravec2006} and its bulk density was estimated to be $2170$ kg/m$^3$, with a $30\%$ uncertainty. 
The size of Dimorphos was estimated to be $164\pm 10$ m, compatible with oblate to prolate shape with axes ratio between 0 and 1.3.
No information on the spin rate of Dimorphos is available, though it is assumed to be synchronous to its orbital period. Separation between components is $1.19\pm 0.03$ km \citep{Naidu2020}. 

Available data do not allow predictions about the internal structure of any of the two bodies. 
Nevertheless, updated estimates of some physical parameters critical to this study available during the publication process of this work \citep{Daly2023} confirm L and LL ordinary chondrites as the best meteorite analogues for Didymos. Considering that typical grain density of such meteorite complexes is in the $3500-3600$ kg/m$^3$ range, the system bulk density is compatible with at least 30\% bulk porosity of its components. Therefore, the primary may  have a gravitational aggregate structure with unknown size distribution of components.

 \subsection{Asteroids on the edge of stability}
 \label{sec:uno}

For some NEAs, the centrifugal force acting on surface particles and boulders at near--equatorial latitudes may slightly overcome the gravitational pull of the asteroid itself in the spinning, non--inertial reference frame of the rotating asteroid. In that case, the radial component of  acceleration for surface particles is directed outwards, allowing them to leave the surface and undergo corresponding dynamical evolution. Leaving the surface does not mean that particles are lost from the asteroid. In fact, they start their motion at zero velocity but non-zero acceleration, and as soon as they lift off they move under the gravitational field of the asteroid, the non-inertial apparent forces due to rotation, the Sun's gravity and its radiation, and --in the case of binary systems-- the gravitational pull of the secondary. Other forces may act as well on small particles on the surface, like electrostatic or molecular forces (cohesion), with the likely result of sticking them together and potentially undergoing the same dynamical effect as dusty clumps. Moreover, small particles, below 1 mm in size, may be lost from the system under the influence of  solar radiation pressure (SRP), but even mm to cm--size particles can have their orbits affected by SRP over a longer span \citep{Yu2017, Ferrari2022, Rossi2022}.
We may expect, instead, that more massive particles potentially levitate for some time, land on the surface and lift off again, repeating such cycles over and over, or just land at latitudes from which further lift off is not possible. Alternatively, they may enter mostly unstable orbits and even transfer to the secondary. 

\cite{Fahnestock2009} studied the effect of particle lofting due to YORP spin up in a binary system, namely 1999 KW4 (Moshup). They found that transferring angular momentum from the primary to the mutual orbit is possible. Regulation of primary spin at the rate for which material lofting takes place may happen so that the orbital angular momentum of the secondary grows steadily. Apparently, lofting occurs in fast transient episodes separated by long periods of slow spin-up. The authors argue that the end state of the system evolution is likely the separation into two asteroids on closely related heliocentric orbits. That may be a potential origin for ``asteroid pairs''. 

\cite{CampoB2013} outlined the possibility of mass lifting as a general mechanism of regolith dispersal in fast spinning Near-Earth asteroids forming  ``dusty'' environment around such bodies.

\cite{Yu2019} investigated mass shedding from the surface of the primary of (65803) Didymos by a semi-analytical approach for shedding conditions. They determined  unstable surface areas by combining the analyses and numerical results of SSDEM
simulations. The authors  find a vast majority of the shedding mass is finally transferred to Dimorphos and leads to a cumulative growth which may cause a spiralling--in orbit of the secondary, an effect going in the opposite direction to  \cite{Fahnestock2009} finding. Further work on failure modes and mass shedding processes was carried out by \cite{Sanchez2016, Zhang2021, Ferrari2022, Hirabayashi2022}.


The main goal of this work is to study the general dynamical features of this mechanism that may be acting on some of the NEA binary systems listed in Table 1. We use asteroid Didymos as a case study, assuming the best available knowledge to the time of developing the model. Obviously, the actual shape, volume and mass of that asteroid --as well as other physical parameters of the system-- will be constrained in much greater detail only after both the DART and Hera missions will characterize the binary system. 
Here we analyze the dynamical evolution of lifted particles as well as their preference for take--off and landing areas on Didymos. We provide mass density distributions of the material that may currently be, or may have been present around it, assuming a given emission rate. 

 The model set up to study particle dynamics in this system and its validation is introduced in Sec. 2; results are presented in Sec. 3 and 4, and  conclusions are discussed in Sec. 5.\\

\section{Model description}
\subsection{Mass loss due to fast spin rate}
To study under what conditions particles may leave the Didymos surface, we considered the available radar-based Didymos shape model, made of 1000 vertices and 1996 facets \citep{Naidu2020} (Future, updated work on this matter shall include the shape model of Didymos available after the DART mission). Test particles were initially assumed to be at rest at the geometric centre of each triangular facet. Particles size distribution follow a differential power law $n(r)\,dn \propto r^{\kappa\,}dr$, with  index $\kappa =  -3.5$. 
We sampled the following particle size range values:
$r_1 = 4.7  \ \mu$m, $r_2=0.1$ mm, $r_3 = 2.3$ mm and $r_4=5.3$ cm. These are central values of the corresponding logarithmic size bins in which the overall size distribution is divided. In this way, we check  particle behaviour from  micron--size,  very sensitive to solar radiation pressure,  to cm--size ---which are typical values for dust grains ejected from active asteroids \citep{Moreno2019, Jewitt2022}--- to multi--cm size, affected only by gravity. Particle density is assumed to be $3500$ $kg/m^3$, according to the L and LL meteorite analogue to the Didymos S spectral type \citep{Dunn2013}.

In the case local acceleration is directed outwards, a particle will take off from the surface and will evolve under the gravitational field of Didymos, the gravitational perturbations generated by the secondary and the Sun, and the solar radiation pressure (SRP), according to the corresponding equation of motion. Under this scheme, at the end of a given integration time, we calculate the total mass of ejected particles in each end state. Based on the trajectories of particles, we have defined four possible end states (ES): ES1, particles that lift off and land again on Didymos’ surface; ES2, particles that remain in orbit; ES3, particles that are accreted onto the secondary; and ES4, particles that escape from the system (see Section \ref{sec:algorithm} for details).
We also estimate the mass density in the Didymos system environment as a function of colatitude, longitude, and distance from the surface of the primary.

\subsection{Equation of motion}
\label{sec:22}
The equation of motion of a particle in a spinning reference system fixed to the primary body can be written as:

\begin{equation}
\begin{aligned}
    & \frac{d^2 \mathbf{r_d}}{dt^2}  = - \nabla \mathbf{U_P} 
    + W_2\frac{\mathbf{r_d}-\mathbf{r_{P \odot}}}{||\mathbf{r_d}-\mathbf{r_{P \odot}}||^3} + W_3 \left[ \frac{\mathbf{r_{P \odot}}-\mathbf{r_d}}{||\mathbf{r_{P \odot}}-\mathbf{r_d}||^3} - \frac{\mathbf{r_{P \odot}}}{r_{P \odot} ^3} \right] +\\
    & +W_4 \left[ \frac{\mathbf{r_{dS}}}{r^3_{dS}}-\frac{\mathbf{r_{PS}}}{r^3_{PS}} \right]   + \boldsymbol{\omega} \times (\mathbf{r_d} \times \boldsymbol{\omega}) + 2 \mathbf{v_d} \times \boldsymbol{\omega}
    \label{Eq:1}
\end{aligned}
\end{equation}

\begin{figure}
\centering
\includegraphics[width=0.9\textwidth]{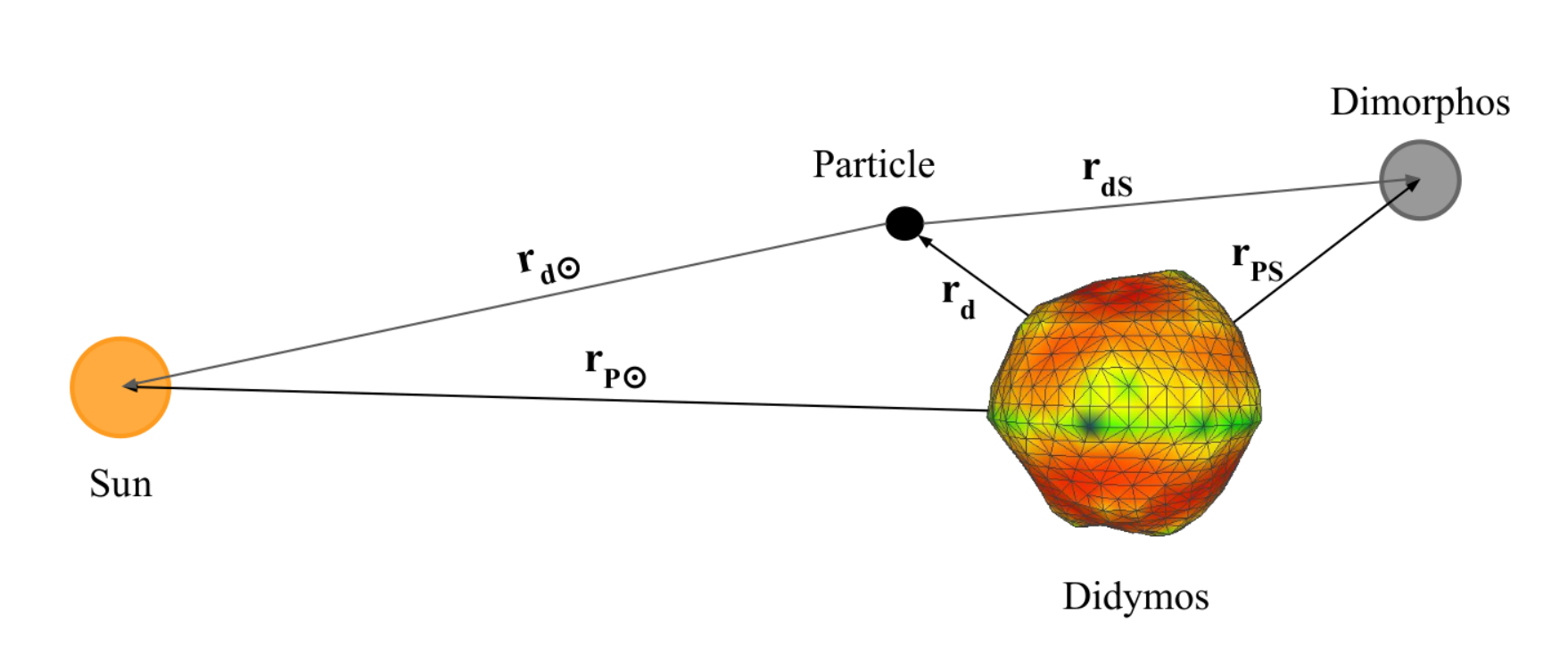}
\caption{Schematic illustration of the system with vectors entering Eq. \ref{Eq:1}.}
\label{fig:fig_esquema}
\end{figure}

\noindent
In that expression, bold characters are vectors, $\boldsymbol{r_d}$ is the primary to particle position vector, $\boldsymbol{r_{P \odot}}$ is the primary to the Sun position vector, $\boldsymbol{r_{dS}}$ is the position vector from the particle to the secondary, and $\boldsymbol{r_{PS}}$ is the primary to the secondary position vector, as shown in Figure \ref{fig:fig_esquema}. 
Here $\boldsymbol{r_{P \odot}}$ $=$ $\boldsymbol{r_d}$ $+$ $\boldsymbol{r_{d \odot}}$, where $\boldsymbol{r_{d \odot}}$ is the vector from the dust grain to the Sun. 
The first term of Eq. \ref{Eq:1} corresponds to the gravitational field per unit mass of the polyhedral shape of the primary, corresponding to the definition in \cite{Werner1994}; 
the second term is the SRP contribution, where 
$W_2 = ({Q_{pr}}/{c})[{E_{\odot} }/({4 \pi})][{\pi d^2}/({4m_d})]$, the following terms are the solar and the secondary gravitational perturbations: $W_3 = GM_{\odot}$ and $W_4= GM_{S}$;  the last two terms are the centripetal and Coriolis force, respectively. The Euler  force can be neglected in this scheme because the spin change is an extremely slow process, lasting orders of magnitude longer than any other change in particle motion. 
Moreover, $G$ is the gravitational constant, $M_P$ is the primary mass, $M_S$ is the secondary mass, $M_{\odot}$ is the mass of the Sun,  $Q_{pr}$ is the efficiency of solar radiation pressure, which is $Q_{pr} \sim 1$ for large absorbing grains \citep{Burns1979}, $c$ is the speed of light, $E_{\odot} = 3.93 \times 10^{26}$ W is the total power radiated by the Sun, $d$ is the particle diameter and $m_d$ its mass: $m_d=\rho_d  (\pi/6)d^3$ , where $\rho_d$ is particle density. 
$\boldsymbol{\omega}$ is the angular velocity of the primary ($\omega=2\pi /T$, where $T$ is the rotational period) and $\boldsymbol{v_d}$ is the velocity of the particle.

\subsection{Description of the algorithm}
\label{sec:algorithm}
Eq. \ref{Eq:1} is integrated numerically  using a fourth-order Runge-Kutta method. 
The primary asteroid can be represented as a sphere or any polyhedral shape. In the first case, the surface of the primary is divided into a grid with $N_{\theta}$ colatitude ($\theta$) bins and $N_{\phi}$ longitude ($\phi$) bins; this grid is extended radially outwards into space with $N_r$ radii ($r$) bins, forming a total of $N_{\theta} \times N_{\phi} \times N_r$ cells. At the beginning, each sample particle is located at the geometrical centre of each surface cell. In the second case, each sample particle is located at the geometrical centre of each triangular facet of the polyhedron, and the $N_{\theta} \times N_{\phi} \times N_r$ space grid is defined outside the body, from the point on the surface farthest from the centre of the body, $\boldsymbol{r_{F}}$. The mass and particle density are computed in those cells. 

To check the validity of the calculation, the primary asteroid was also modelled as a sphere, and the gravitational field of the corresponding homogeneous sphere was analytically calculated. In the general case, the gravitational field corresponding to the polyhedral model is computed following the guidelines given by \cite{Werner1994} at each point in space. In both cases, a point-mass 
secondary body is assumed. Dimorphos follows a circular orbit  on the equatorial plane of Didymos. Moreover, the position and orientation of the binary in the heliocentric ecliptic system are rigorously computed at each time step from its available ephemeris and system pole orientation. The position of the Sun, as seen from the primary, is also computed, and the shadow of the primary on the ejected particles is checked at each time step to set the SRP on each particle either on or off. The shadow produced by Dimorphos on Didymos is not taken into account, as it is less than $5\%$ of the Didymos surface, smaller than other sources of uncertainty in the model. Particles are initially at rest on the surface. A detachment condition is applied to check whether any given particle has an outwards component of local acceleration, 
which depends strongly on the apparent centrifugal force in the rotating system of the primary. Detachment occurs when $\sum_{i} \boldsymbol{F_i} \cdot \boldsymbol{n} > 0$, where $\boldsymbol{F}_i$ stands for each force corresponding to the accelerations in Eq. \ref{Eq:1}, and $\boldsymbol{n}$ is the unit vector normal to the surface facet
and directed outwards.
Particles, once ejected from the surface of the primary, move in the gravitational field  of the two bodies plus the radiation pressure and gravitational perturbing forces of the Sun. \\

Based on the trajectories of the particles, we define four final states,  as follows:
\begin{itemize}
    \item ES1, landing particles: particles with radial distance from the centre is equal to or less than $r_F$, the particle is checked to be outside or inside the surface defined by the shape model. In the first case, the integration process continues, otherwise, the algorithm finds the intersection between the trajectory of the particle and the facet of the shape model. The collision coordinates are recorded, and the particle is labelled as ES1.
    \item ES2, orbiting particles: particles belong to this group if at the end of the integration time they are still in orbit.
    \item ES3, particles accreted on Dimorphos: during numerical integration, the position of a given particle  with respect to Dimorphos is checked at every time step. The gravitational field of Dimorphos is considered as a point--mass source. The case in which the distance between the particle and the position of Dimorphos is less than its equivalent radius,  is considered as a collision, and the particle is labelled as ES3.
    \item ES4, escaping particles: particles located beyond $10^4$ m from the centre of mass of Didymos belong to this group. 
    Even if the Hill's radius of the system is $75$ km, the limit is set at $10$ km  
    distance from the centre of the system for practical reasons. This is a distance beyond which we found that only a negligible mass density contribution is missed from rare particles orbiting back from outer distance. In fact, mass density, even at the Dimorphos distance, is extremely small, and the density profile keeps decreasing further away. 
    This has the benefit of hugely reducing  both the storage load and the computational cost with no effect on the mass density calculation.
\end{itemize}

\begin{figure}
\centering
\includegraphics[width=0.8\textwidth]{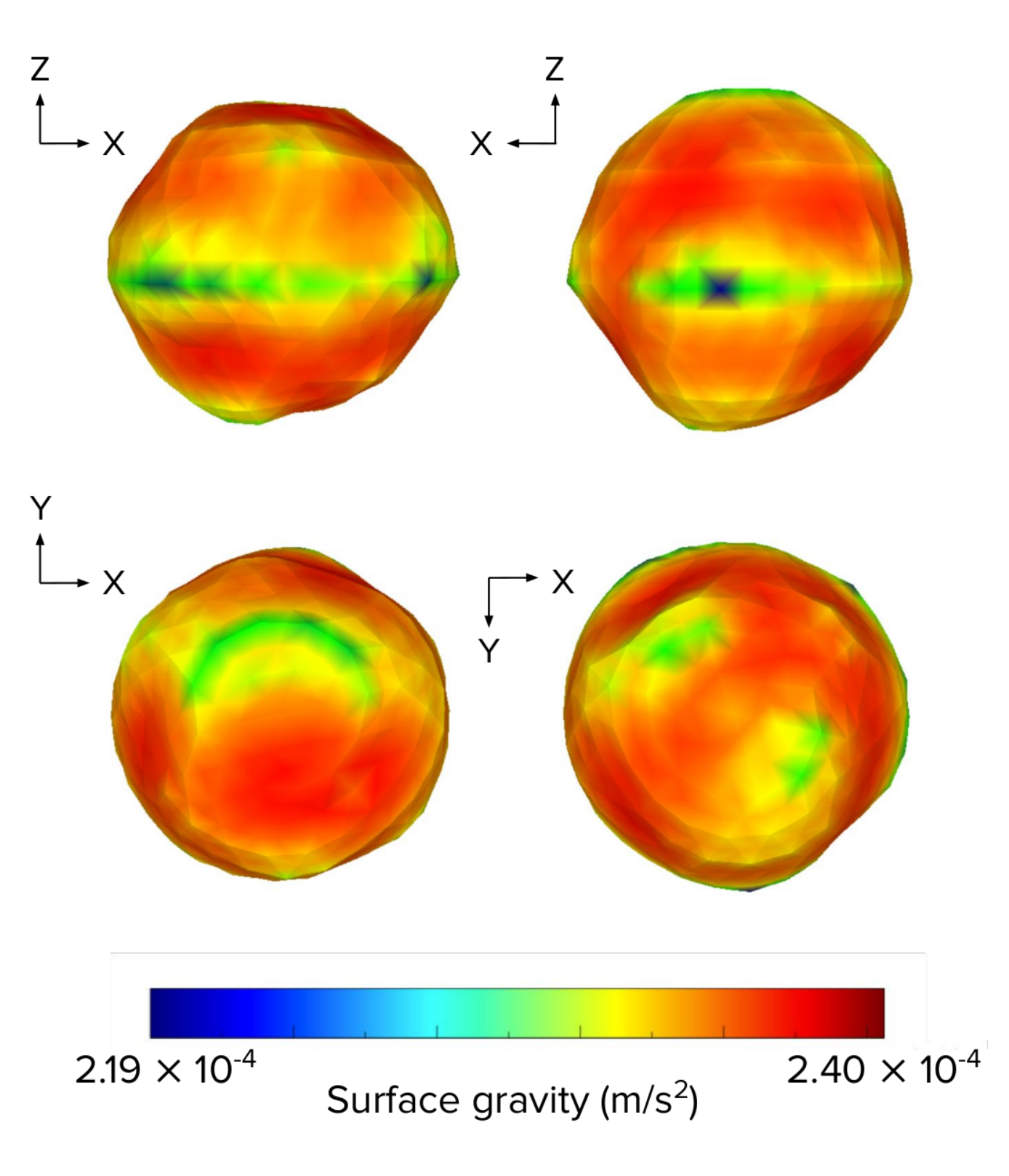}
\caption{Surface gravity on a non--rotating Didymos. Top left and right: view of both hemispheres. Bottom left and right: north and south polar region view, respectively.}
\label{fig:surf_gravity}
\end{figure}

 Figure \ref{fig:surf_gravity} shows the surface gravity map of the Didymos shape model of a non-rotating Didymos, built using a mass of $5.229 \times 10^{11}$ kg with homogeneous density. This is worked out considering that the overall size of Dimorphos, $D_{S}\approx 0.2 \, D_{P}$, so that the relationship between masses is close to $M_{S}/M_{P} \sim 0.01$. Lowest gravity regions are clearly located on the equatorial bulge (blue-green colour in Figure \ref{fig:surf_gravity}). \\
We assessed the difference between the gravity field of Didymos, taking into account its shape model, against a sphere of equal mass and density, by calculating the radial gravitational field. Obviously, differences are found close to the surface (until approximately 570 m), but both fields  converge at larger distances, as expected. 

In our model, the difference between the spherical and the polyhedral shape of Didymos is in the definition of spatial cells. Given that the shape model represents an irregular surface, it is not possible to simply define a surface grid and extend it into space in a uniform way. Instead, we set the longitude and latitude grid  starting at the vertex of the triangular facet at the maximum distance to the centre  ($r_{F}$). For the Didymos shape model, this spherical inner grid surface starts at $r_{F}=427$ m, which is the smallest distance at which mass density outside the body is calculated. This is the minimum radial distance from which spatial density computation is performed.  In other words, this is the spherical surface from which the computation of the mass density is made outside Didymos. The drawback of this approach is that no information is available inside the tiny region in between the shape model surface and the beginning of the 3D grid. 

As for the modelling of the shadow cone produced by Didymos, given the  moderate depart of the asteroid shape from sphericity, for the sake of simplicity a profile corresponding to a sphere of radius $r_{F}$ is assumed instead of the shape model itself. That approach has the advantage to save  computing time, with a negligible impact on the volume of the projected shadow cone. In addition, in order to save CPU time, the gravity field of the primary is approximated by its corresponding spherical field at $r > 550$ m, given the proximity of spherical and polyhedral solutions at that distance. The relative error is less than $3\%$ at that point, decreasing at higher values of r.

\begin{figure}
\centering
\includegraphics[width=0.65\textwidth]{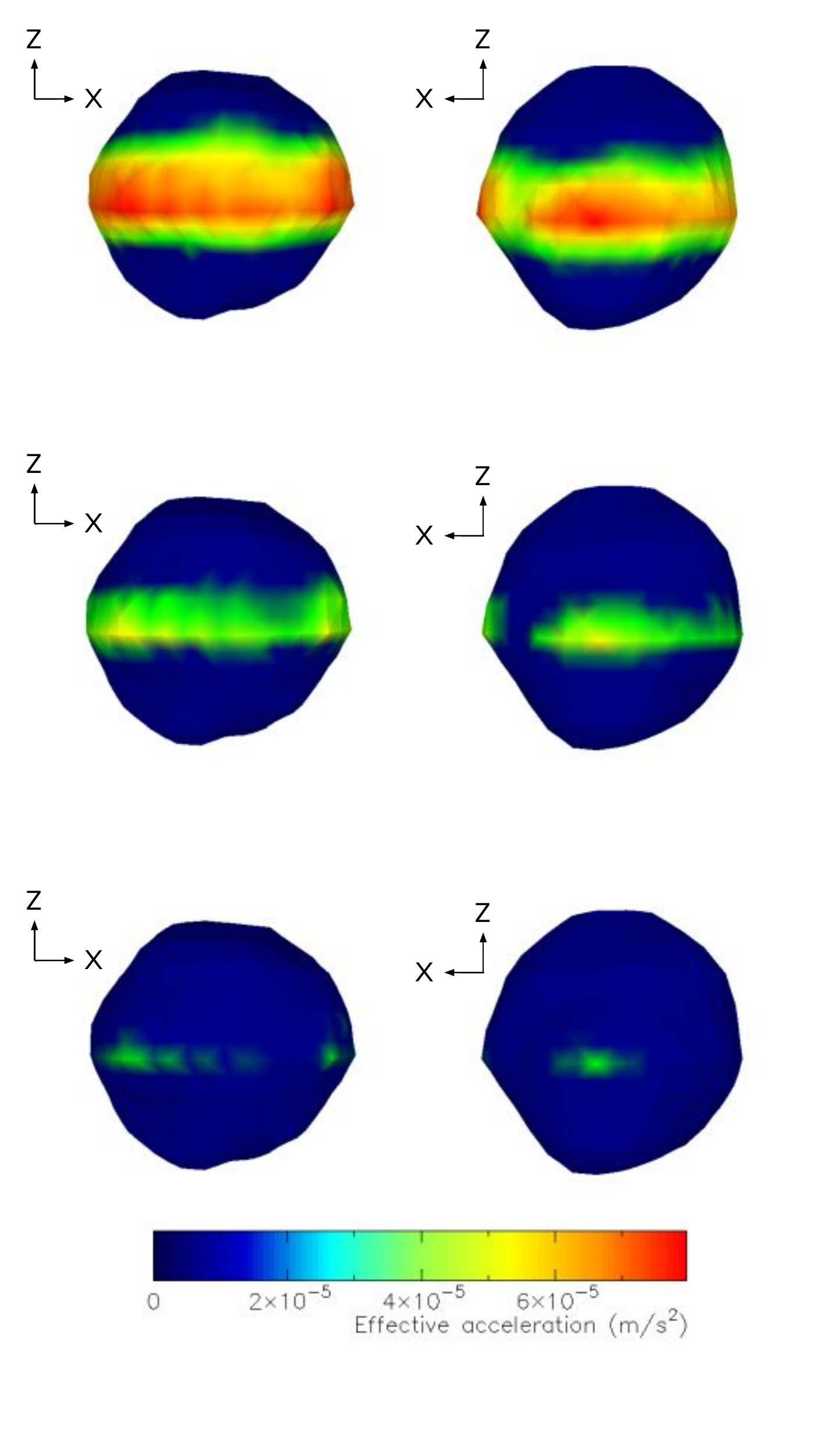}
\caption{Effective acceleration on Didymos surface (gravity plus apparent centrifugal force) for different Didymos bulk densities. Only acceleration with non-nil outward radial component is plotted. Top: low density case with 1813 $kg/m^3$, same physical parameters as in simulation ($M_2$, $V_6$). Mid: nominal density case, 2104 $kg/m^3$, same physical parameters as in simulation ($M_4$, $V_4$). Bottom: high density case, 2411 $kg/m^3$, physical parameters as in simulation ($M_7$, $V_3$). The corresponding ratios of centripetal
to gravitational acceleration at average equatorial radii ($a_c/g=\omega^2r_{eq}^3/(GM_P)$) are 1.29,
1.11, and 0.97, respectively. Notice that even in the ($M_7$ , $V_3$) case, with average $a_c/g < 1$,
equatorial surface irregularities may allow for lift-off locations.}

\label{fig:local_gravity}
\end{figure}

The code input data are the orbital and physical parameters of the system extracted from the mentioned Hera Didymos Reference Model (ESA internal document). Further inputs are the perihelion epoch, the dates of start and end of integration, the integration time step used in the Runge-Kutta procedure, the mass loss rate, and  particles  properties, i.e.,
particles radii, their density and the exponent of their size frequency distribution (SFD). 
Other running parameters are also given: the number of particle size bins, which are conveniently spaced logarithmically to be consistent with the power-law distribution, the time bins, 
the radius limit for mass density computation, and the escaping distance, i.e., the distance at which particles are considered to escape the system and no longer contribute to density in the considered space range. See Sec. \ref{sec:time} and \ref{sec:mass_calculation} for details.

\subsubsection{Time integration procedure}
\label{sec:time}

When we look at the image of a comet tail, we need to keep in mind that particles were ejected from the comet surface at any time before the time at which the image was taken, let's call that time $t \leq T_{obs}$.
A similar situation can be envisioned in the case of  Didymos ejected particles.
The following numerical procedure for reproducing such a situation is applied. 

Let's call $T_{start}$ the time at which the program checks, for the first time, the particle surface detachment condition at any given surface facet. The final $T_{obs}$ time is instead the time at which the calculation of the particle mass and number density is made (in the analogy above, this corresponds to the time at which the observation of the system environment is made). The total integration time is $T_{obs}-T_{start}$, which is divided by the number of time intervals
$n_t$, so that 
$\Delta t=({T_{obs}-T_{start}})/n_t$ is the length of each time integration period. One sample particle of each size is located on every surface facet. At the beginning ($T_{start}$),  the detachment condition for each particle size is checked, if this is met, the sample particle starts its motion under the action of the corresponding forces by a $4th$--order R-K integrator, with an integration time step $H$. This is done for all particle sizes. 
Many different values for the R-K integration time step were tried, the value finally adopted corresponds to the largest one safely allowing for convergence in the integration.
Once integration is over, that is, at $t=T_{obs}$, the same procedure is repeated at the initial time $T_{start} + \Delta t$ ($\Delta t>>H$), and so on.
This is necessary because we need to consider the evolution of particles potentially leaving their facet  at all $n_t$ equally spaced times between $T_{start}$ and $T_{obs}$.
The same process is carried out for all surface facets. 
In this way, the emission of sample particles is checked along the whole integration time. At time $T_{obs}$, the number of particles in orbit (ES2) is counted at each space cell. Instead, particles that hit the surface of the primary or the secondary (ES1, ES3), or reach the escape distance (ES4) at some time, do not contribute to  space density (ES2), but their end state is recorded.
At the end of the whole integration time for all surface facets, the data corresponding to the end--state of all sample particles are stored, as well as their position in space and mass density as a function of longitude, colatitude and radial distance from the centre of Didymos.

\begin{figure}
\centering
\includegraphics[width=0.9\textwidth]{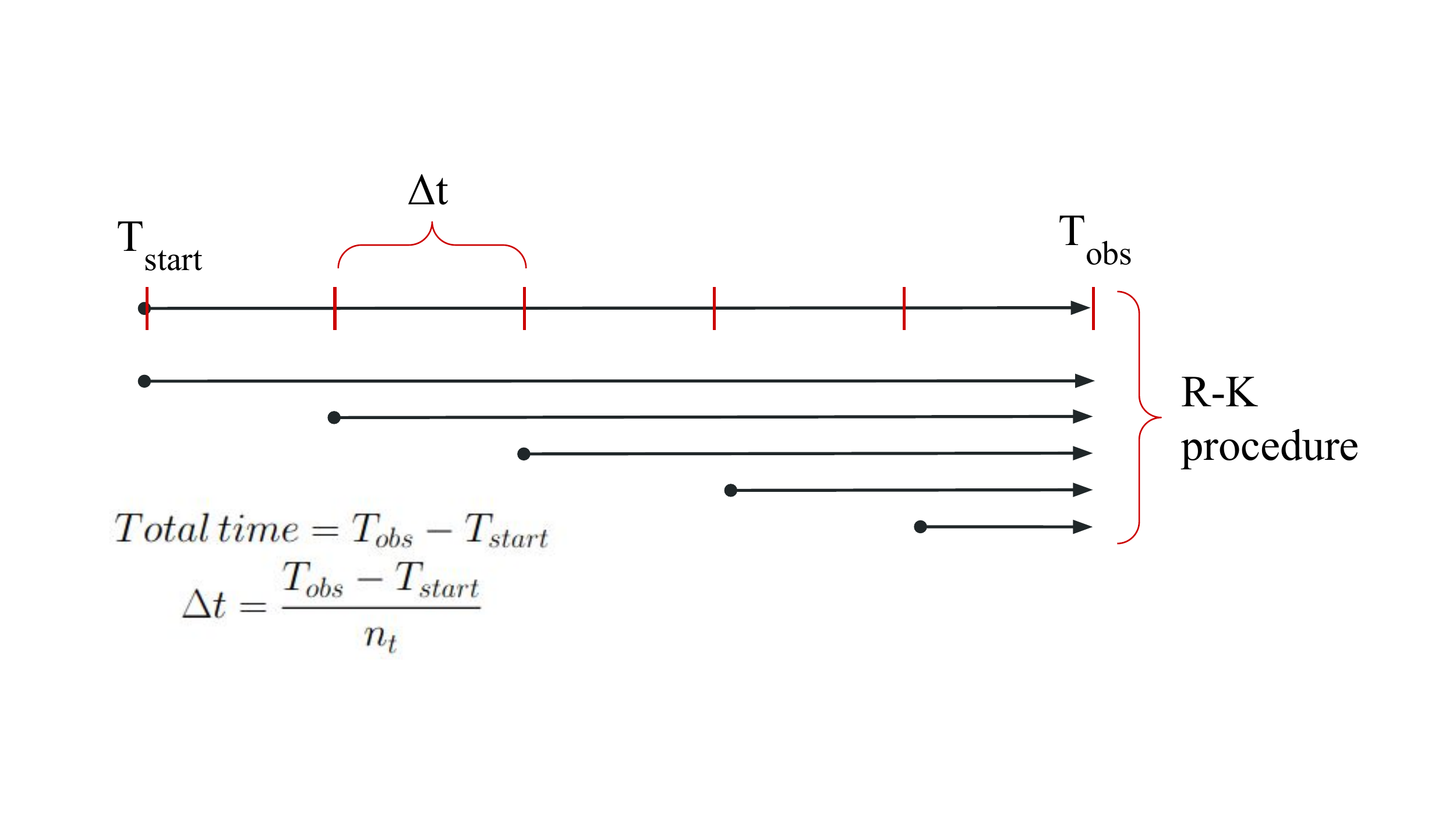}
\caption{Integration time scheme for a sample particle initially at any surface cell or facet. Each particle is integrated $n_t$ times, starting at different times spaced by $\Delta t$. The corresponding  Sun-Primary-Secondary orbit configuration is updated accordingly at all $n_t$ times.}
\label{fig:fig_time_bin}
\end{figure}

\subsubsection{Mass density calculation}
\label{sec:mass_calculation}

Once the end state of sample particles is known, it is necessary to calculate the absolute mass in each end state. Therefore, we face the problem of how to translate sample particle statistics into the corresponding actual mass  in each end state.
First of all, we need to make assumptions on the mass emission rate; in addition, the mass and number of particles have to be scaled with the adopted SFD.

Active asteroids show a diversity of mass--loss mechanisms including sublimation, impacts, fast rotation, electrostatics, thermal effects, etc. Estimates of rotation mass shedding in fast spinning rubble-pile asteroids range from $10^{-7}$ kg/s, reported for asteroid (101955) Bennu \citep{Lauretta2019a, Lauretta2019b, Hergenrother2019}, to $2.3$ kg/s \citep{Hui2019}, and 35 kg/s \citep{Jewitt2019a, Sanchez2019} in the case of (6478) Gault. Episodic mass losses ranging from 1 to 10 kg/s have been derived for multi--tailed asteroid P/2013 P5 (PANSTARRS) \citep{Moreno2014}, which can be clearly attributed to mass shedding due to rotational instability \citep{Jewitt2015b}.
To date, no observation of this kind of activity has been reported for Didymos, so this parameter is unknown. For this reason, we considered a constant reference value of mass production rate for Didymos, that was set arbitrarily to $dM/dt = 1$ kg/s for the whole asteroid. A suitable fraction of such value is used only for cells for which the lift off condition is met, no particle will be ejected  for most surface cells far from the equator though. 
The contribution of ejected particles to the mass  and number density  in each space cell is computed as follows. The mass ejected per unit surface area is calculated as 
$(dM/dt)/(4\pi R_p^2)$, where $dM/dt$ is the assumed mass production rate on the whole surface, $R_p$ is the Didymos radius, and the mass ejected per surface cell in a time interval $\Delta t^{\prime}$ is:

\begin{equation}
M_{cell}=\frac{1}{4\pi R_p^2}\frac{dM}{dt} S_{cell}\Delta t^{\prime}
\end{equation}

where $S_{cell}$ is the area of the surface cell.  The total mass is distributed according to the already mentioned SFD in the radius range $[r_{min}$, $r_{max}]$.
In a given  size bin $[r_i$, $r_j]$ within such range,  the number of particles emitted from each cell in a given time interval is: 

\begin{equation}
    N_{bin}= M_{cell} \frac{3}{4}  \frac{1}{\pi \rho_{particle}} \frac{\int_{r_i}^{r_j} r^\kappa dr}{ \int_{r_{min}}^{r_{max}} r^{\kappa+3} dr}
\end{equation}

and the corresponding emitted mass is: 

\begin{equation}
    M_{bin}=N_{bin} \frac{4}{3} \pi \rho_{particle} \frac{\int_{r_i}^{r_j} r^{\kappa+3} dr}{  \int_{r_i}^{r_j} r^{\kappa} dr}
\end{equation}

The mass and the number of particles are updated accordingly in every space cell where sample particles are located at the end of  integration. The mass and number of particles are finally divided by the cell volume to get the corresponding mass and number density.

\subsection{Model validation}

\begin{figure}
\centering
\includegraphics[width=0.8\textwidth]{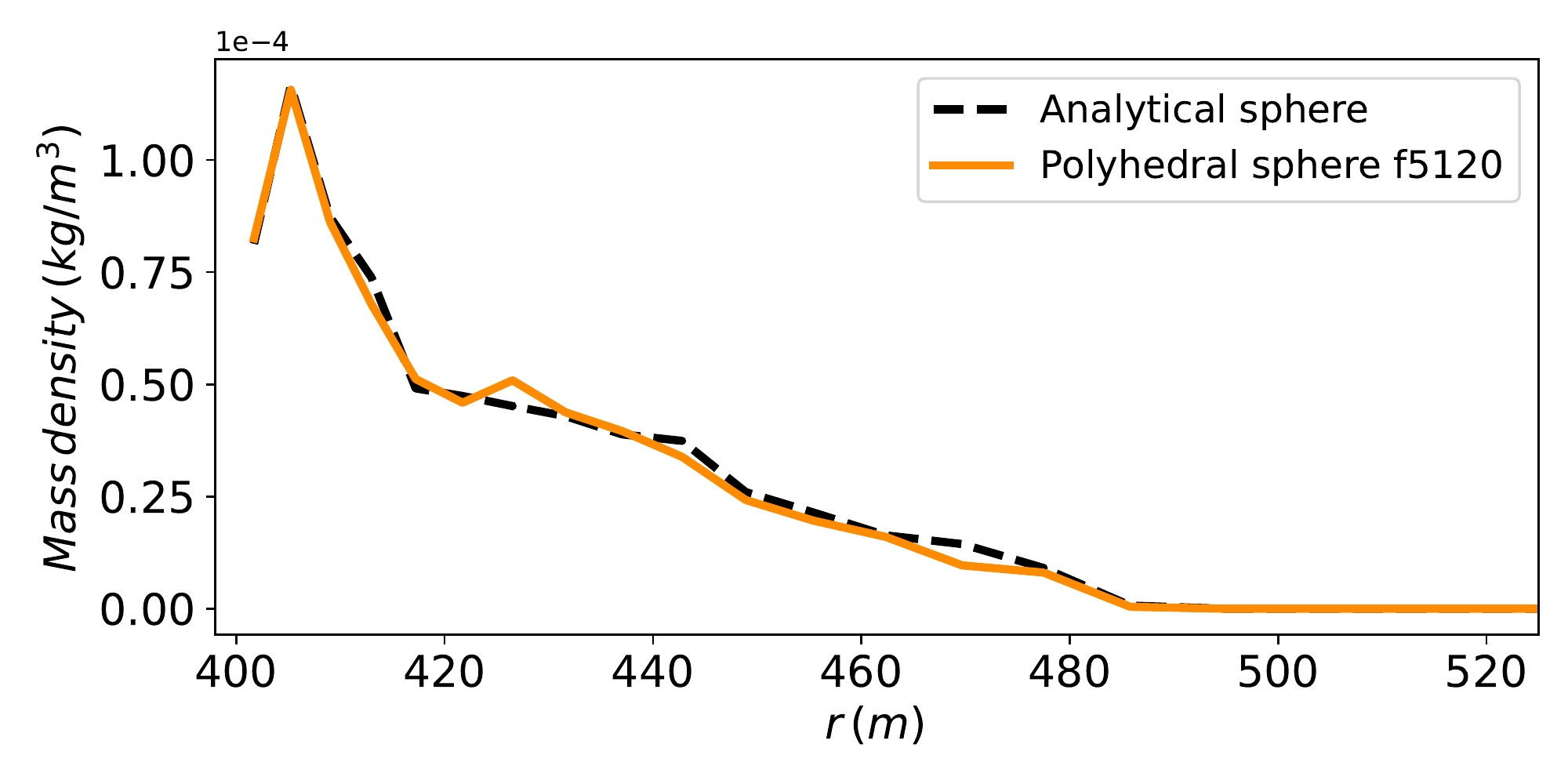}
\caption{Comparison between radial mass density distributions for the analytical and polyhedral (5120 facets) representation of a sphere.}
\label{fig:fig_test}
\end{figure}

In order to check the validity of our model when applied to the polyhedral representation of solid bodies, we compared the results obtained for the dynamical evolution of particles departing from analytical and polyhedral representations of a spherical body made of 5120 facets (see Figure \ref{fig:fig_test}).
In both cases, the same mass and equivalent size of Didymos were used, $M_P=5.229\times10^{11}$ kg and $R=400$ m, respectively. The density profiles are in very good agreement even at small distances from the body surface. Simulations were run over 30 days. \\

\section{Summary of model simulations}

The current Didymos system physical parameters are known with  wide uncertainties (see Sec. 1).   
For this reason, we first mapped the mass--volume parameter space and related each pair of values with the corresponding total orbiting mass obtained by our model. Then, we focused on the nominal parameters for the Didymos mass and size and we performed a detailed analysis of  particle  detachment and landing process by means of numerical simulations. The wide heliocentric distance of the Didymos system ---due to its high eccentricity--- led us to study the particle behaviour around very different locations (perihelion and aphelion) of the system, and under the corresponding initial epoch conditions, in the case of full orbit integrations.
The whole set of simulations carried out is outlined next.

\begin{itemize}
    \item Simulation 1: Conditions for mass lift-off. 
    
    \noindent
    The first goal is to study under what conditions it is possible to find mass around the primary. Thus, we combined 7 values of each mass and volume parameters of Didymos within estimated uncertainty. That resulted in 49 numerical runs, one per each corresponding value of  bulk density (see Table \ref{tab:tab_s1}). We let the system evolve during 30 days, near the perihelion epoch, starting on August 28, 2020 and ending on September 26, 2020. Performing simulation runs around perihelion is the worst case for particle survival, as SRP is more efficient in perturbing their motion and taking particles away from the system.
   
   \item Simulation 2: Evolution in one full orbit (starting at perihelion). 
    
    \noindent
    Here we used the nominal mass and volume of Didymos in order to set its nominal bulk density. The simulation was run over a full heliocentric orbit of Didymos (770 days), starting and ending around the perihelion epoch (from August 28, 2020 to October 7, 2022). 
    The behaviour of small ($r_1$, $r_2$ and $r_3$) and large ($r_4$) particles were analyzed separately to study the effect of the SRP.

    \item Simulation 3: Evolution in one full orbit (starting at aphelion).
    
    \noindent
    We used the same physical parameters as in Simulation 2, but the simulation was run over a full heliocentric orbit of Didymos (770 days), starting and ending around the aphelion epoch (from September 15, 2021 to October 25, 2023). Again, the analysis of small and large particles was done separately. 
    
    \item Simulation 4: Short term evolution around perihelion.
    
    \noindent
    Here we report the results of Simulation 1 again for the nominal  mass and volume of Didymos primary. The system evolved over 30 days around the perihelion epoch, starting on August 28, 2020 and ending on September 26, 2020.

    \item Simulation 5: Short term evolution around aphelion. 
    \noindent
    Analogous to Simulation 4, but over 30 days near the aphelion epoch, starting on September 15, 2021 and ending on October 15, 2021.
    \end{itemize}
    
We run short term (30 days) simulations to be able to catch the main features of the  orbiting mass for the whole set of 49  runs in Simulation 1 and to underline (Simulations 4 and 5) the dependence of the outcome for small particles on the observation epoch. 
This choice, in the case of Simulation 1, was motivated, on one hand,  by the fact that full orbit simulations are very time consuming. On the other hand,   the behaviour of particles ---as far as mass distribution is concerned--- does not change around a given epoch, beyond some 20  days. Therefore, we considered a safe strategy to perform 30--day runs around the least favorable epoch for particle survival.

Orbital parameters used in  simulations are shown in  Table 2. Table 3 summarizes the volume and mass parameter space ranges, and the system physical parameters are introduced in Table 4. 

\begin{table}[t]
\begin{center}
\begin{tabular}{ l  c  c }
\hline
\multicolumn{3}{ c }{Didymos system} \\ \hline
Semimajor axis & $1.6444327821$ & au \\
Eccentricity & $0.38393203178$ & - \\
Inclination & $3.40808504153$ & deg \\
Argument of perihelion & $319.245071345$ & deg \\ 
Longitude of ascending node & $73.2392391311$ & deg \\
Perihelion epoch & $11.6146$/$9$/$2020$ & dd/mm/yyyy \\ \hline
\multicolumn{3}{ c }{Dimorphos secondary orbit} \\ \hline
Semimajor axis & $1190$ & m \\
Orbital period & $11.9216$ & h \\ \hline

\end{tabular}
\caption{Orbital parameters of the Didymos system used in the simulations. (Source: Didymos Reference Model, ESA)}
\label{tab:tab1}
\end{center}
\end{table}

\begin{table}[t]
\begin{center}
\begin{tabular}{ c  c  c  c  c  c  c }
\hline
\multicolumn{7}{ c }{Mass ($\times 10^{11}$ kg)} \\ \hline
$M_1$ & $M_2$ & $M_3$ & $M_4$ & $M_5$ & $M_6$ & $M_7$ \\ 
4.687 & 4.867 & 5.048 & 5.229 & 5.409 & 5.590 & 5.771 \\ \hline
\multicolumn{7}{ c }{Volume ($\times 10^{8}$ m$^3$)} \\ \hline
$V_1$ & $V_2$ & $V_3$ & $V_4$ & $V_5$ & $V_6$ & $V_7$ \\  
 2.209 & 2.301 & 2.393 & 2.485 & 2.584 & 2.684 & 2.783 \\ \hline

\end{tabular}
\caption{Input values in Simulation 1. Different Didymos bulk densities are obtained combining mass and volume values.}
\label{tab:tab_s1}
\end{center}
\end{table}

\begin{table}[t]
\begin{center}
\begin{tabular}{ l  c  c }
\hline
Parameter & Value & Units \\ \hline
Didymos rotation period & $2.26$ & h \\
Didymos mass & $5.229 \times 10^{11}$ & kg \\
Didymos volume & $2.48548175 \times 10^8$ & m$^3$ \\
Didymos density & $2104$ &  kg/m$^3$ \\ 
Dimorphos diameter & $164.0$ & m \\
$N_r\times N_{\theta} \times N_{\phi}$ & 50 x 36 x 36 & - \\
$\Delta t$ & $10.0$ & min \\
Time integration step ($H$) & $100$ & s \\
Distance for mass density computation & [$r_F$, 1500] & m \\
Escaping distance for ES4  & $1 \times 10^4$ & m \\ \hline
\end{tabular}
\caption{Set of input parameter values for the Didymos system in Simulation 2 to 5. Nominal mass and volume correspond to the ($M_4$, $V_4$) set in Table \ref{tab:tab_s1}}
\label{tab:tab2}
\end{center}
\end{table}

\section{Results}

\subsection{Conditions for mass lift-off (Simulation 1)}

\begin{figure}
\centering
\includegraphics[width=0.8\textwidth]{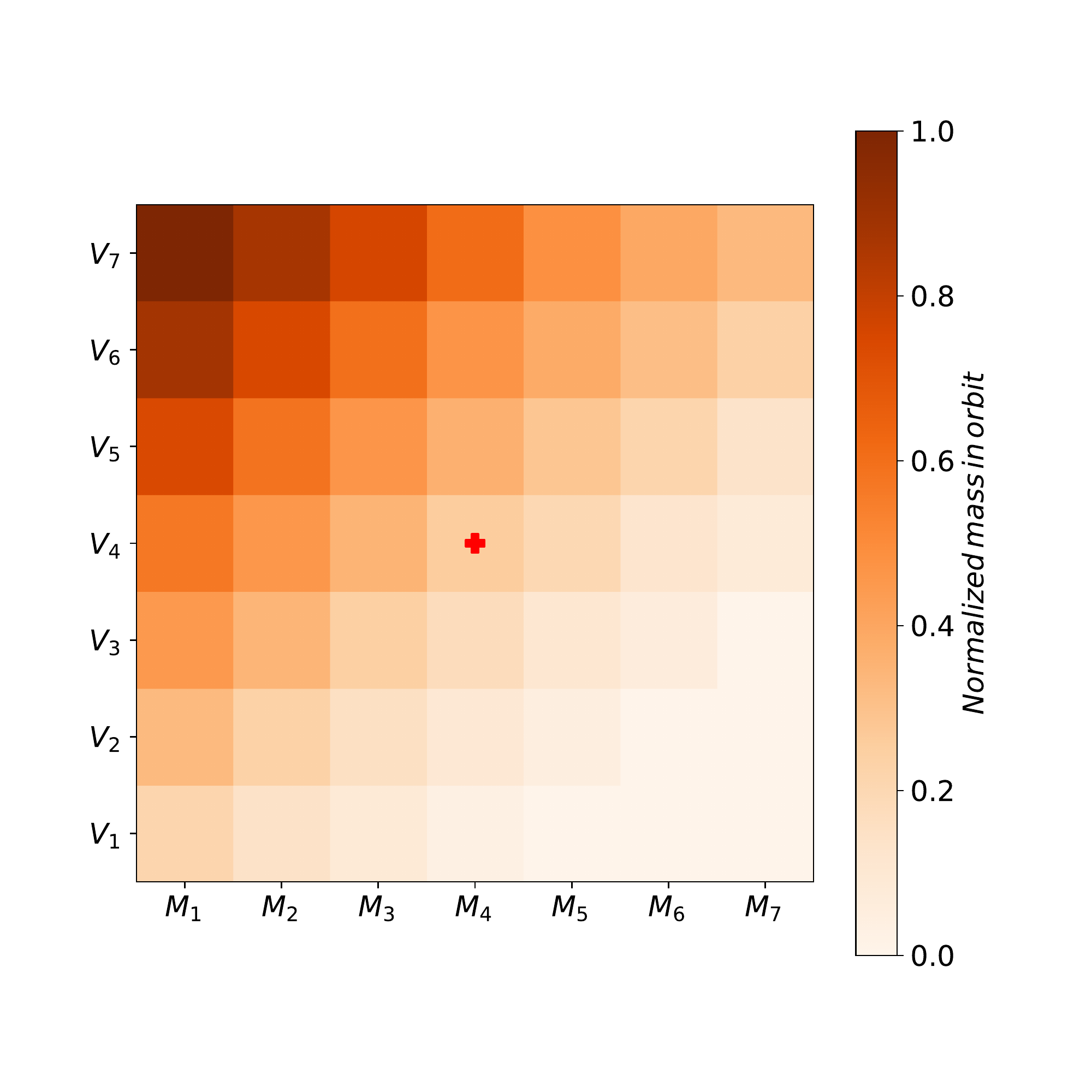}
\caption{Results of Simulation 1. 
$M_i$ and $V_j$ are input values for the modelled mass and volume of Didymos. Each square represents the spatial mass corresponding to different $(M_i,\ V_j)$ sets (see Table \ref{tab:tab_s1}). The colour bar shows the output orbiting mass normalized to the maximum value, in the case the evolution is followed during 30 days around perihelion. The red cross stands for the nominal  bulk density. Simulations corresponding to $(M_6,\ V_1)$, $(M_7,\ V_1)$ and $(M_7,\ V_2)$ result in no mass in orbit at perihelion.}
\label{fig:fig_s1}
\end{figure}

In Figure \ref{fig:fig_s1}, each square gives the orbiting mass at the end of each simulation after 30 days around perihelion, normalized to the maximum orbiting mass value. The central cross corresponds to the nominal case, which will be analyzed further in the next sections (Simulations 2 to 5). Particle detachment takes place at the nominal spin rate for most values in the chosen Didymos physical parameter space, except for three parameter combinations (see Table \ref{tab:tab_s1}). 


\subsection{Particle evolution after one full orbit (Simulations 2 and 3)}

In Simulation 2 and 3, we analyze particle end states after they leave the surface and the corresponding lifetimes (Table \ref{tab:fates}).
Depending on size, more than $97  \%$   of particles fall back onto the surface of the primary (ES1). The time required to land back also depends on particle size. In general, for any ES, the larger the particle, the longer its lifetime. Particles of radius $r_1$ spend on average $0.81$ h orbiting the asteroid before landing on its surface, while particles $r_4$ may orbit more than 4 hours before landing. In addition, as expected, most  particles that escape  the system (ES4) correspond to size $r_1$. However, that end state probability is only $2.81\%$. When such particles pass from the shadow cone generated by the asteroid to the illuminated region, they are quickly removed from the system, typically after $6$ hours. On the contrary, a small fraction of $r_2$, $r_3$ and $r_4$ size particles are eliminated, and they are able to survive in the environment of the asteroid between 37 and 600 hours, before they are removed. Instead, particles that are accreted onto the secondary (ES3) are mostly $r_4$, followed by $r_3$. Such particles stay in orbit for some 100 and 81 hours, respectively, before reaching Dimorphos. $r_1$ and $r_2$  particles have lower probabilities of reaching the moonlet, taking 2.5 to 4 hours, driven by SRP. At the end of the integration time, there are mostly large particles orbiting the system. $r_3$ size particles are able to stay in orbit about 46 h in Simulation 2 and 74 hours in Simulation 3, and this difference is  related to still some weak interaction with SRP
at such size range. On the other hand, $r_4$ particles have a median lifetime in orbit of $\sim 100$ hours and no influence of the SRP is detectable any longer.

These results can be easily understood by comparing the effect of the gravitational force ($F_g$) due to Didymos on any given  particle, with respect to the corresponding solar radiation force ($F_{rad}$), as a function of the distance to the centre of the asteroid ($a$):

\begin{equation}
\label{Eq:2}
    \frac{F_g}{F_{rad}}=\frac{4}{3} \frac{G  M_P c \rho_d  \frac{d}{2}}{Q_{pr} F_{\odot}} \frac{1}{a^2}
\end{equation}

\begin{figure}
\centering
\includegraphics[width=0.8\textwidth]{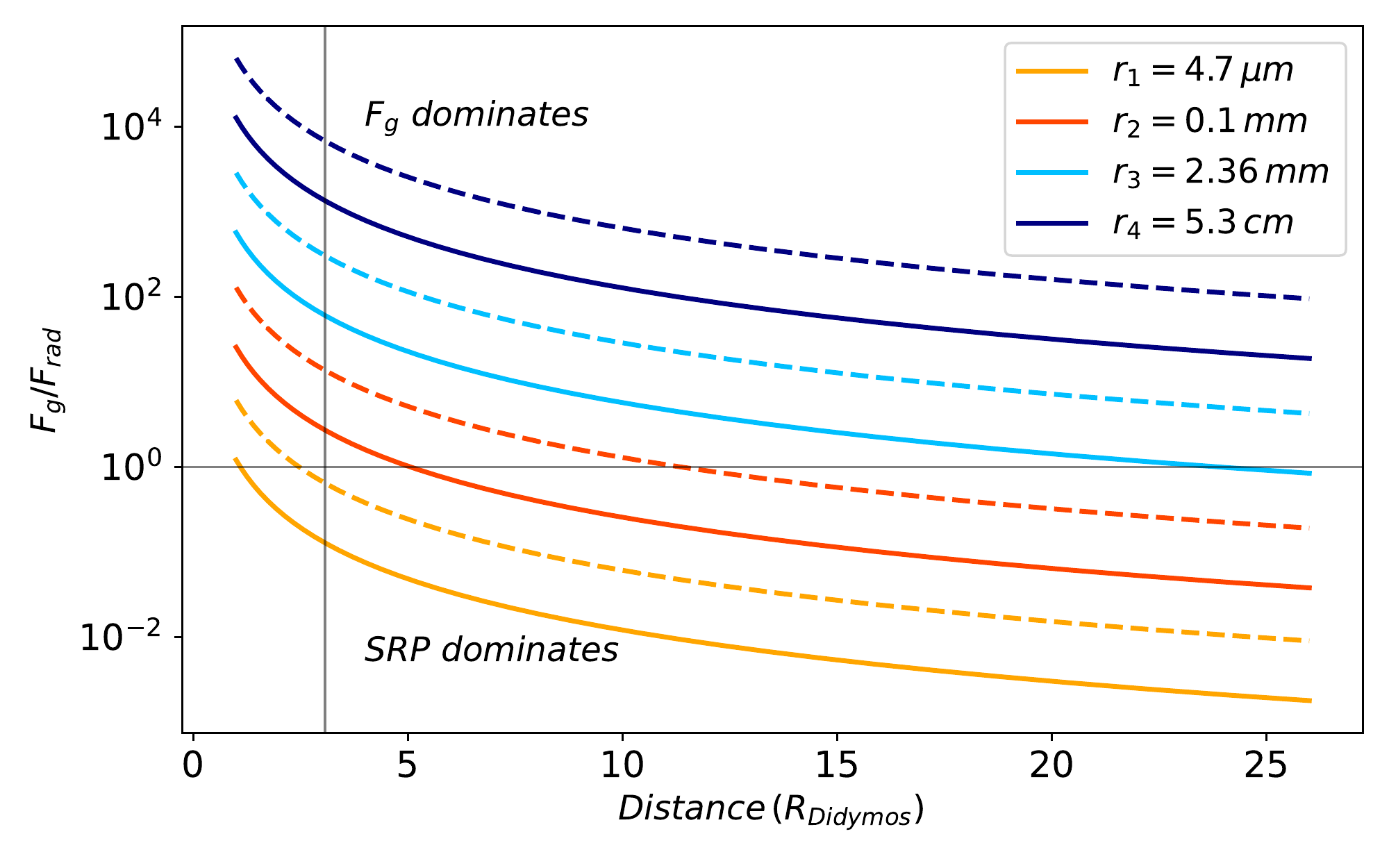}
\caption{Ratio between the gravitational force of Didymos ($F_g$) and the solar radiation force ($F_{rad}$) acting on  particles, as a function of the distance to the centre of the asteroid. The horizontal gray line divides space into two regimes, one is dominated by gravity and the other by SRP  force. The gray vertical line indicates the distance to Dimorphos. Coloured curves represent different particle sizes. Solid and dashed curves correspond to motion around perihelion and  aphelion, respectively.}
\label{fig:fg_vs_srp}
\end{figure}

$F_{\odot} = E_{\odot} / 4 \pi a^2$ is the solar flux at the heliocentric distance of the asteroid system, and the other parameters are the same as in Section \ref{sec:22}. 
The ratio between the two force contributions is plotted in Figure \ref{fig:fg_vs_srp}. A  horizontal gray line separates the distance to Didymos  into two regimes, one is dominated by gravity and the other one by solar radiation force. The gray vertical line shows the distance to Dimorphos and the coloured curves represent the relationship between the two mentioned forces for different particle  sizes. Solid and dashed curves correspond to motion around  perihelion and  aphelion, respectively. Particles of size $r_1$ evolve almost completely under the action of SRP, whereas particles of size $r_2$ are affected by both SRP and $F_g$. Instead, motion of particles of size $r_3$ and $r_4$ are dominated by the force of gravity. That implies that  there is no smooth transition between the behaviour of the particles of size $r_1$ and $r_3$ and the corresponding mean lifetimes in ES2.
Lifetimes are shorter around perihelion (Simulation 2) than around aphelion (Simulation 3) due to different SRP force values. These results are comparable with the work by \cite{Ferrari2022}. They also show that, in such low-gravity environment, SRP plays an important role in the dynamics of small dust grains. Instead, particles larger than a few millimetres are mostly affected by the Didymos gravitational force within the orbit of Dimorphos.

\begin{landscape}
\begin{table}[H]
\begin{center}
    \begin{tabular}{ c  c | c  c | c c | c c | c c }
    \hline
 & & ES1($\%$) & $t_l$ (h) & ES2($\%$) & $t_l$ (h) & ES3($\%$) & $t_l$ (h)  & ES4($\%$) & $t_l$ (h) \\ \cline{1-6}
\hline
\multirow{4}{2.5 cm}{Simulation 2} 
 & $r_1$ & $97.114$  & $0.81$ & $0.002$  &$3.55$& $0.070$  &$2.61$& $2.814$ &  $6.28$ \\
 & $r_2$ & $99.975$  & $3.14$ & $0.009$   & $2.65$ & $0.014$  & $3.94$& $0.002$ & $37.2$ \\
 & $r_3$ & $98.682$  & $3.83$ & $0.066$   & $45.5$& $1.190$ &$81.4$& $0.062$  & $152$ \\ 
 & $r_4$ & $97.383$  & $4.44$ & $0.181$   & $99.9$& $2.206$ &$102$& $0.230$  & $604$ \\   \hline
\multirow{4}{2.5 cm}{Simulation 3} 
 & $r_1$ & $97.120$ & $0.81$ & $0.001$  & $2.37$ &  $0.071$ & $2.58$ & $2.808$  & $6.28$ \\
 & $r_2$ & $99.964$ & $3.14$ & $0.021$  & $5.57$ &  $0.013$ & $3.92$ & $0.002$  & $38.6$ \\
 & $r_3$ & $98.628$ & $3.83$ & $0.134$  & $73.9$ &  $1.175$ & $81.4$ & $0.063$  & $147$ \\ 
 & $r_4$ & $97.376$ & $4.47$ & $0.184$  & $102$  &  $2.214$ & $101$  & $0.226$  & $586$ \\  \hline
\end{tabular}
\caption{Percentage of particles in each end--state (ES) and the corresponding median life time ($t_l$). $r_1= 4.7 \mu m$, $r_2= 0.1$ mm, $r_3= 2.36$ mm and $r_4= 5.3$ cm.}

\label{tab:fates}
\end{center}
\end{table}
\end{landscape}

\subsection{Density distribution of orbiting mass}
\subsubsection{Small particles}
\label{sec:small}
\begin{figure}
\centering
\includegraphics[width=0.8\textwidth]{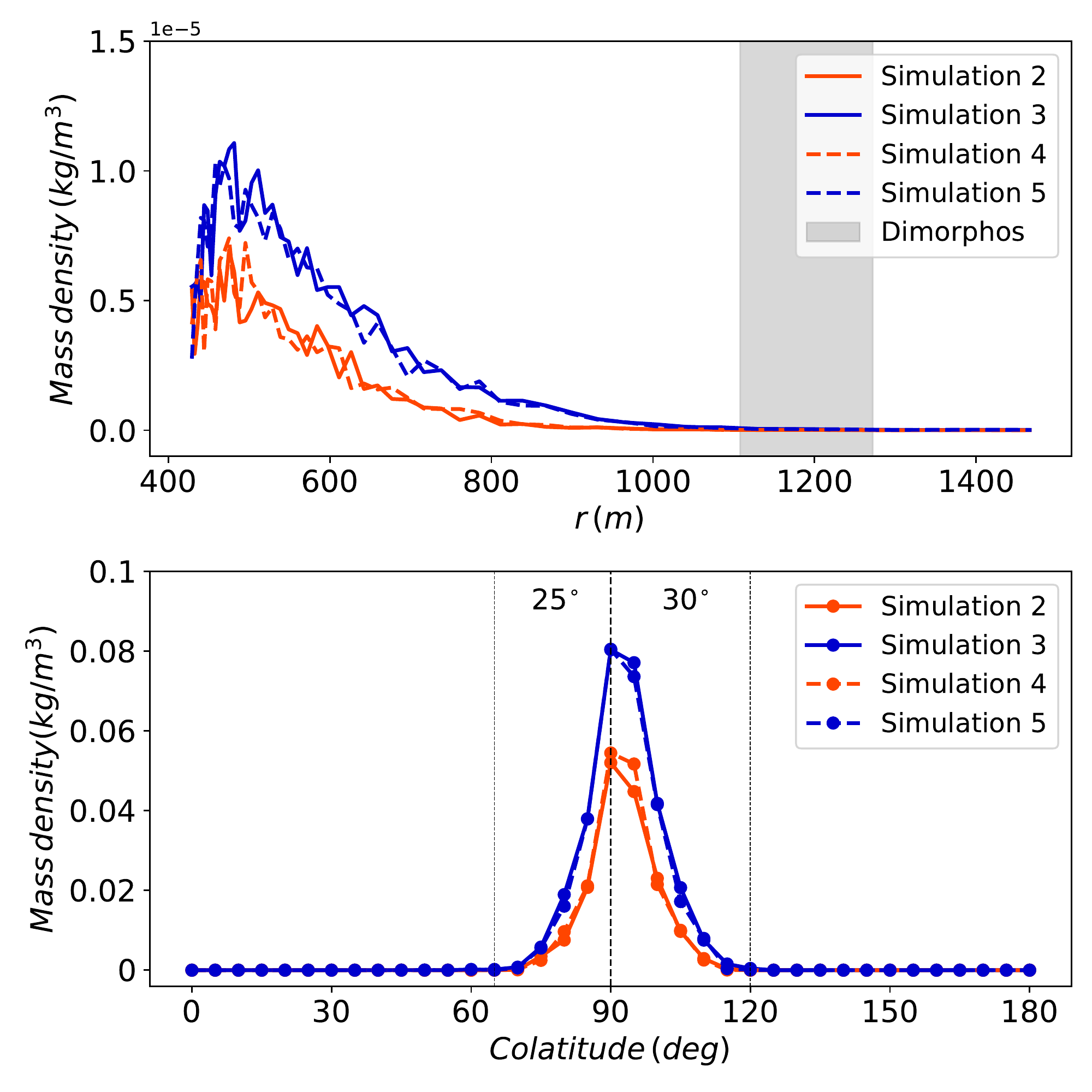}
\caption{Density profiles of mass in orbit around Didymos corresponding to small particles: $r_1$, $r_2$ and $r_3$. Top: radial profile after one full heliocentric orbit (Simulations 2 and 3), and after 30 days around perihelion and aphelion epochs (Simulations 4 and 5). Bottom: mass density profile as a function of colatitude for Simulations 2, 3, 4 and 5. Simulations outcome correspond to the assumed overall value for mass ejection rate ($1$ kg/s).}
\label{fig:fig_s23}
\end{figure}

Here, we focus on the analysis of the behaviour of small particles ($r_1$, $r_2$ and $r_3$) that are in ES2 at the end of Simulations 2, 3, 4 and 5.
Mass density profiles as a function of radial distance from the asteroid centre are shown in Figure \ref{fig:fig_s23} (top). This is obtained by integrating on each cell in colatitude and longitude at any given radial distance. Plots begin at $r_{F}=427$ m, the origin of spatial cells. 
In density profiles, maxima are found at $r\sim475$ m in the case of Simulations 2 and 4, at $r\sim481$ m in Simulation 3, and at $r\sim458$ m in Simulation 5. Therefore, the largest mass density of orbiting particles is  31 to 54 meters above the mentioned reference distance from the centre $r_{F}$.
Beyond such distance, mass density steadily decreases, as expected. As a result, orbiting mass density in simulation  runs around perihelion (Simulation 4) is smaller than for the corresponding simulations around aphelion (Simulation 5).
This is due to the strong influence of SRP on small particles.
The outcome of density profiles for 30 days integration time are very similar to those obtained integrating over a full orbit of the system around the Sun. This is the case both for perihelion and aphelion epoch starting time. Therefore, the amount of small particles in the system environment  at a given time  strongly depends on the epoch at which the mass density observation is made.

The colatitude density profile (See Figure \ref{fig:fig_s23}, bottom), together with the radial density profile, show that orbiting particles are mostly located in a thin disk, with maximum density on the equatorial plane. However, the disk is not symmetrical at north and south of the equator due to Didymos topographical shape model  inhomogeneities. On the northern hemisphere, landing location is spread over 25 degrees, between $65^{\circ}$ and $90^{\circ}$ of colatitude, while, on the southern hemisphere, the distribution extends up to 30 degrees, between $90^{\circ}$ and $120^{\circ}$ colatitude. The density mass difference between  observation  about perihelion and aphelion is also shown in the colatitude profile (Figure \ref{fig:fig_s23}, bottom). 

The analysis of particle motion and mass density shows that although most  ejected particles return to the surface of the primary with short orbit lifetimes (a few hours), a continuous lift off process supplies them into space.

\subsubsection{Large particles}
\begin{figure}
\centering
\includegraphics[width=0.8\textwidth]{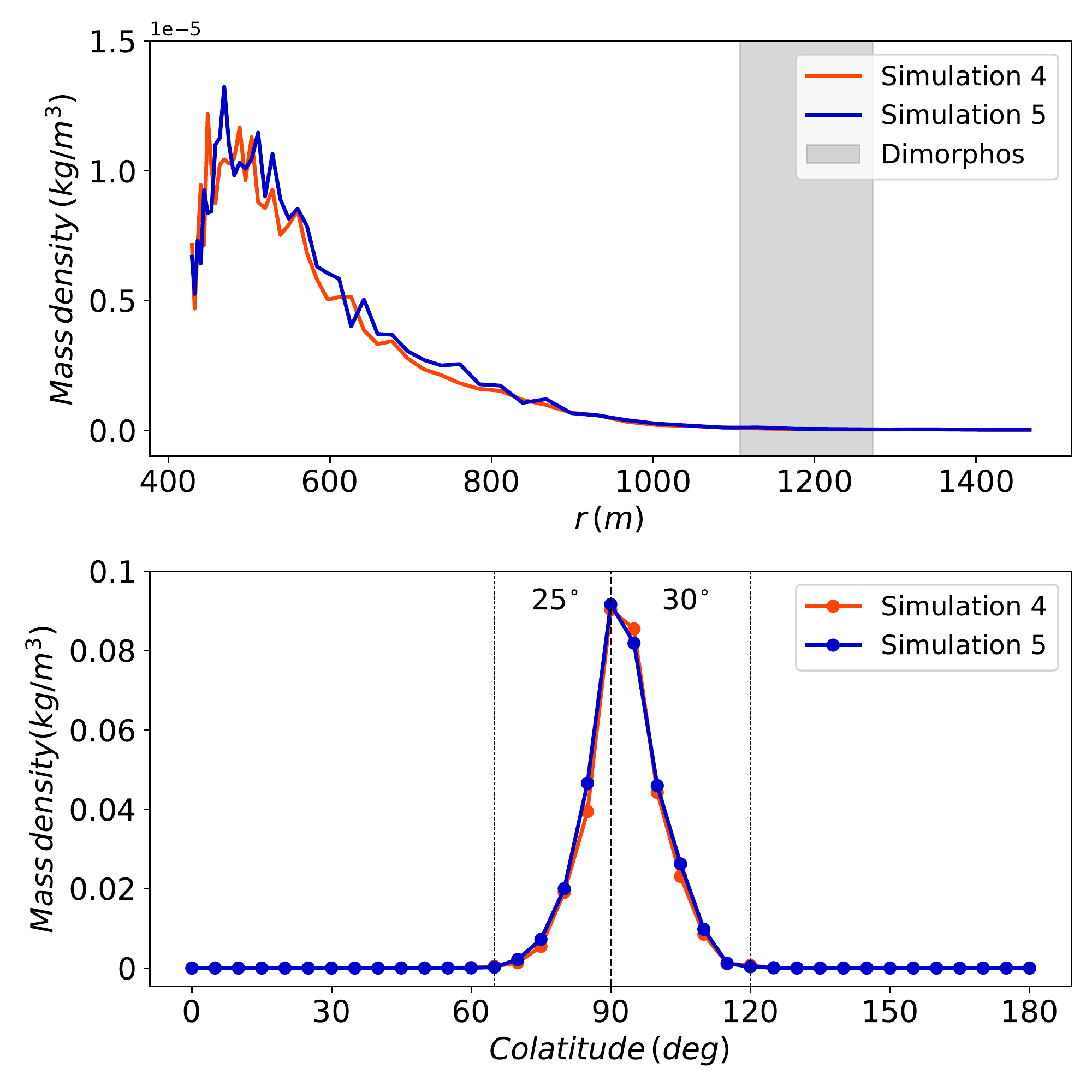}
\caption{Density profiles of mass in orbit around Didymos, including small particles, $r_1$, $r_2$, $r_3$, and large particles, $r_4$. Top: radial profile after 30 days around perihelion (Simulations 4) and aphelion (Simulations 5). Bottom: mass density profile regarding colatitude.}
\label{fig:large2}
\end{figure}

This section extends the analysis of results including particles of radius $r_4$, not affected by SRP. Figure \ref{fig:large2} shows the corresponding radial and colatitude mass distributions. This also  applies to any object of size larger than $r_4$, as they are only affected by gravitational and non--inertial force terms.

\begin{figure}
\centering
\includegraphics[width=0.8\textwidth]{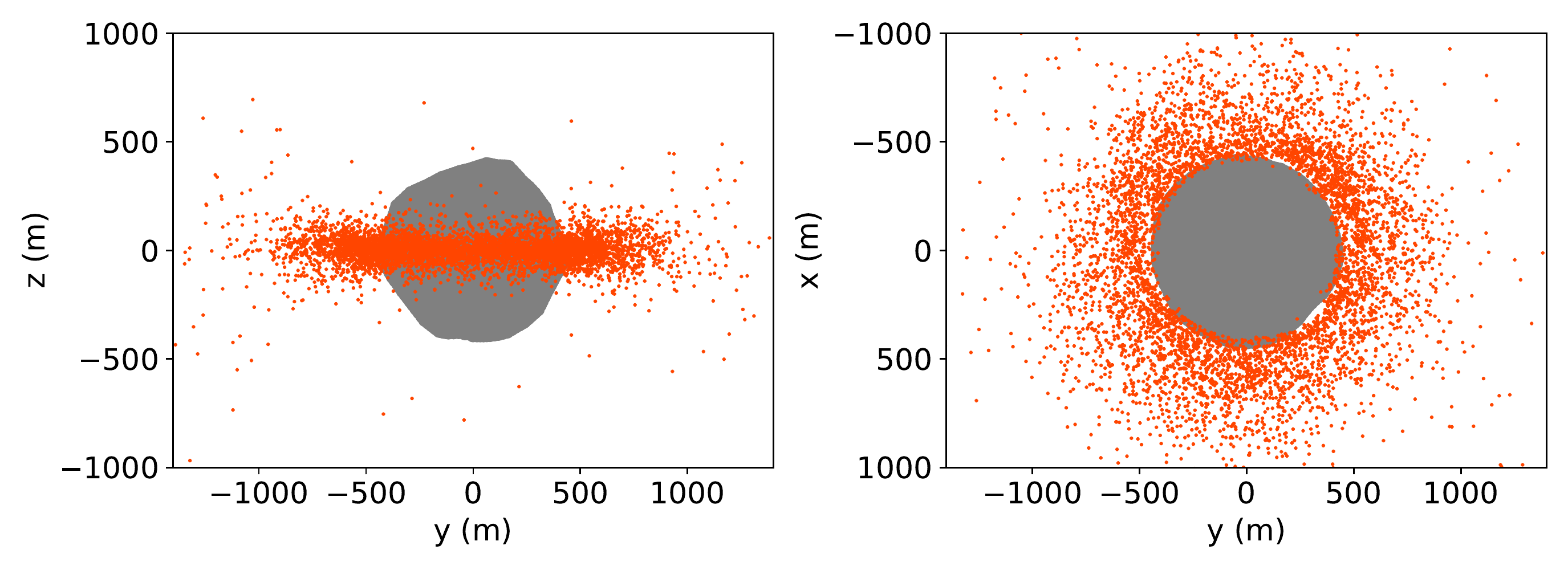}
\caption{Side and polar view snapshot of orbiting particles around Didymos at the end of Simulation 4. Particles of all sizes are included. All particles are represented by not-to-scale dots.}
\label{fig:fig_orbiting}
\end{figure}

When large particles are considered, there is no  substantial difference in orbital mass density at aphelion with respect to perihelion, although the radial density profile at aphelion (Simulation 5) is slightly larger than the corresponding density at perihelion (Simulation 4) over the whole considered domain. Large particles are the greatest contributors to orbiting mass density, which is equal around both epochs. Instead, small particles contribute more at aphelion than perihelion, building up the very small difference between densities observed around the two different epochs (see Figure \ref{fig:large2}).
 The disk--like distribution of particles in orbit at a given time is shown in Figure \ref{fig:fig_orbiting}, in the case of Simulation 4.\\

It is interesting to notice that the probability ($p_2(r)$) for small particles to have an orbital end state (ES2)  is smaller as size decreases (see Table 5). Similarly, duration of orbits is shorter for small particles with respect to large ones ($\delta t_2(r)$, normalized). Both effects combine in the size frequency distribution of particles found to be in orbit at  a given time, that can be described as $n_o(r)dn= n (r)dn \, p_2(r)\, \delta t_2(r)\, dr\propto r^{-3.5}\, p_2(r)\, \delta t_2(r)\, dr \propto r^{{\kappa}_o}\, dr$, with ${\kappa}_o\approx -3$. In summary, the SFD of particles    observed   in orbit (ES2) looks to be shallower than the original asteroid surface population assumption (${\kappa}= -3.5)$, and is therefore skewed towards large  particles.

\subsection{Distribution of landing particles}
\begin{figure}
\centering
\includegraphics[width=1\textwidth]{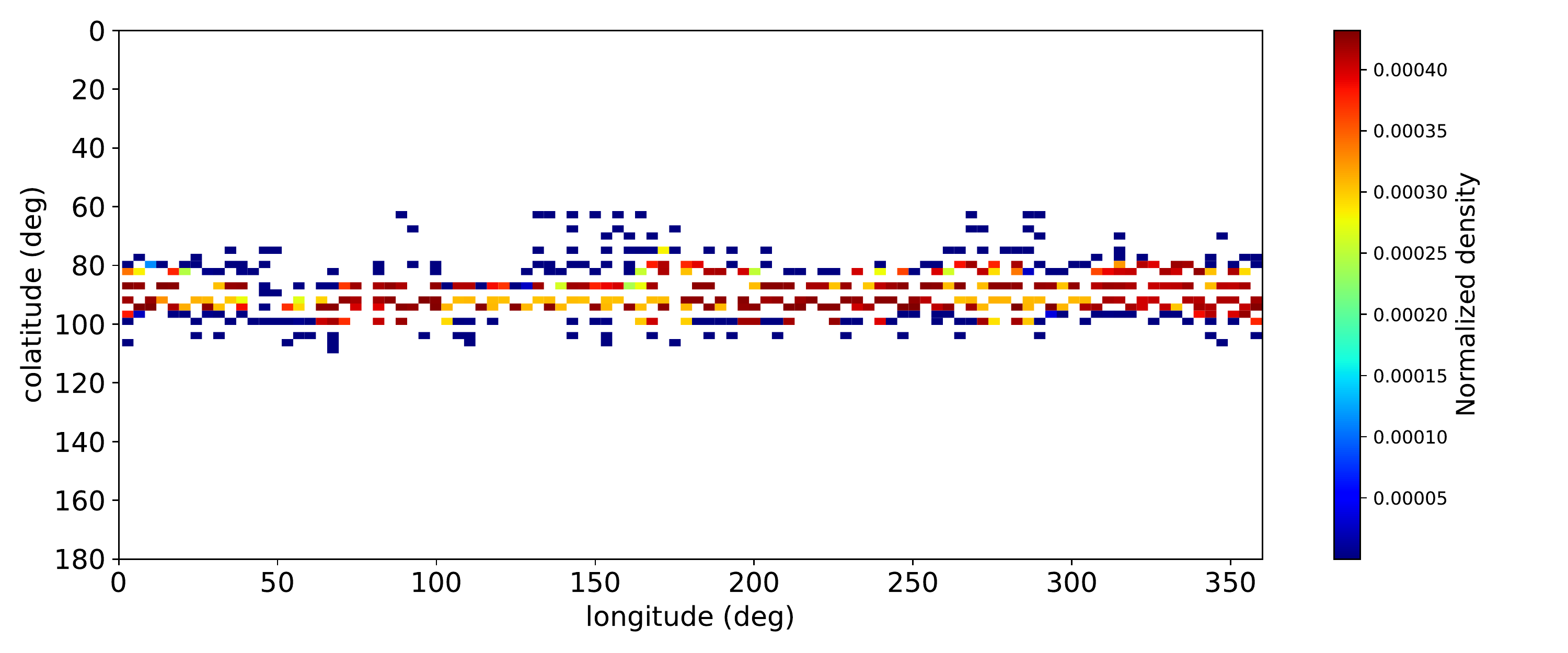}
\caption{Initial positions of lifted-particles on Didymos' surface. At the beginning, the position corresponds to the geometrical centre of the triangular facets. The plot corresponds to Simulation 3.}
\label{fig:pos_salida}
\end{figure}

\begin{figure}
\centering
\includegraphics[width=1\textwidth]{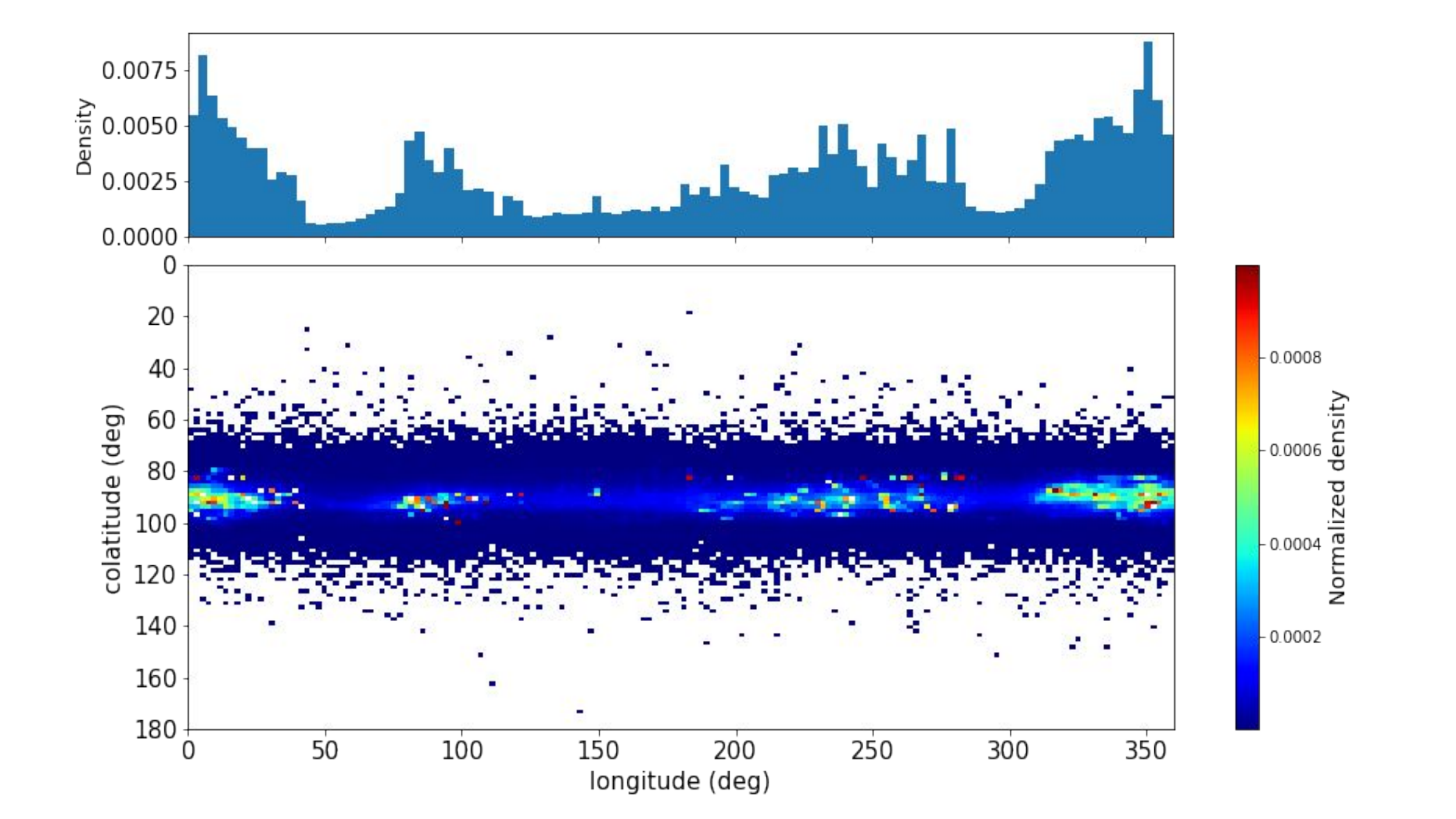}
\caption{Final particle landing sites on Didymos surface. Top: landing distribution in longitude. Bottom: colour map with the landing positions in longitude and colatitude. The plot corresponds to Simulation 3.}
\label{fig:pos_caida}
\end{figure}

Figure \ref{fig:pos_salida} shows ---for Simulation 5, as an example--- a mass density colour map distribution of the initial position of particles that are able to lift off the surface. 
At the beginning of any run, particles lie at the geometrical centre of the shape model triangular facets; hence a discrete distribution shows up in Figure \ref{fig:pos_salida}. 
The detachment area surrounds the equatorial plane, but it is not symmetrical with respect to it. The maximum detachment latitudes to the north and to the south of the equator are $\theta_N \sim 28 ^{\circ}$ and $\theta_S \sim 19 ^{\circ}$, respectively, in all the cases, in Simulations 2 to 5. After detachment, most lifted particles (more than $97\%$) land back on the surface of the primary. Figure \ref{fig:pos_caida} (bottom) shows landing locations after 30 days of evolution.
During this time, the gravitational field of Didymos, the gravitational perturbation of Sun and Dimorphos, and SRP, modify particle trajectories, increasing their orbital inclination and allowing them to reach mid and high landing  latitudes, typically higher in comparison with the detachment positions. The colour map shows that the density distribution peak in colatitude lies near the equatorial plane. It is important to note that particles  landing at low latitudes can likely leave the surface again. Indeed, such  particles are a reservoir for later material ejection. Instead, particles falling at mid and high latitudes would not be dynamically able to lift off again from their new location. 

However, mass sliding mechanism towards low latitudes are known to have taken place on asteroids Bennu and Ryugu \citep{Sabuwala2021}, such effect may drive mass back to the equatorial region, re--fuelling the take off and landing process. Figure \ref{fig:pos_caida} (top) shows  non-homogeneous landing distribution in longitude. This behaviour is related to the topography of the asteroid in the equatorial region. Particles flight heights are low, so lifted-particles preferentially land on relatively high terrains around the equator.
\\

\section{Discussion and conclusions}

Primary fast spin up in NEA binary systems is relatively common. We study the possibility that mass is able to take off the surface of such bodies, taking (65803) Didymos NEA binary system as a study case because it is the DART (NASA) and Hera (ESA) space missions target.
The latest available shape model for the asteroid at the time of carrying on this study was used.
We set up a model to study the dynamical behaviour of the particles that are able to take off the surface  due to the primary fast spin. The gravitational field of the asteroid is taken into account, together with all non--inertial forces in the asteroid rotating frame. Perturbations due to the satellite, Dimorphos, and to the Sun are suitably considered, as well as  SRP. 

We explored the Didymos primary mass and volume parameter space to assess under what asteroid density conditions particle detachment from the surface is possible. We find that most values in that parameter space  allow for particle  lift off. In the nominal case, that is possible in a region limited to the $(+28^{\circ}, -19^{\circ})$ latitude interval around the equatorial plane, non--symmetric due to the primary irregular shape.
Available range of mass and volume estimates is compatible with the possibility of having a disc of particles of different sizes around Didymos. The actual mass density is hard to predict as the rate of mass emission is unknown. Nevertheless, our model finds the distributions in colatitude and distance from the asteroid according to assumed parameters. Maximum values for the mass density distribution show up at very low latitudes and at some $30$ to $50$ m away from the average primary equatorial radius, decreasing steadily at larger distance.

Most  lifted particles ($>97\% $) do land back on the surface, with variable lifetimes depending on size, but typically smaller than 5 h. Small particles have shorter lifetimes than large ones because of  SRP. In fact, when small particles take off on the day side of the asteroid, they are quickly pushed back to the surface, and their flight time is shorter the stronger the SRP is. When such particles take off in the night side, they follow ballistic trajectories until they typically land back or, in the case they reach the terminator, they are pushed away from the system by SRP.
Particle landing locations are not evenly distributed in longitude, concentrating instead in equatorial morphology highs. This is because   most particles only reach low altitudes with respect to their original take off location, and they eventually stumble into such highs on their flight path. It is interesting to notice that the latitude landing range distribution is wider than that of lift off distribution location. As a consequence, some particles may land at latitudes from which they can no longer  take off.

SRP has an important effect on small particle dynamics, to the point that, for such particles, mass density in space happens to be a function of the orbital phase at which it is observed ($T_{obs}$). 
Didymos has a very eccentric orbit, such that heliocentric distance changes from 1.013 to 2.276 au. This results in very different contributions from the solar tide and SRP at different distance  from the Sun. Total mass surrounding Didymos at aphelion is more than double ($\approx$ 9/4) than at perihelion for sub--cm size particles. 
Instead, looking at the overall mass in orbit, including large particles ---not affected by SRP--- no significant difference is found, due to the contribution of large particles to total mass. The size distribution of particles in orbit should be shallower (${\kappa}\approx -3$) than their size distribution on the asteroid (${\kappa}=-3.5$) due to smaller probability and lifetime in orbit of small particles with respect to large ones.

In conclusion, our model shows that the presence of mass in orbit around   Didymos, or similarly shaped fast spinning NEAs, is possible. That includes the potential presence of boulders around the system, that may eventually  reach the secondary ($\approx 2\%$ probability) at this stage.
In any case, as outlined in Section \ref{sec:uno}, the existence and  observability of mass around the system may depend on which primary rotational phase and which secondary orbital phase Didymos is undergoing at the observation time, according to the mechanism described by \cite{Fahnestock2009}. 

Right after the completion of this study, the DART spacecraft successfully impacted Dimorphos, and first results derived from the images of both the DRACO camera on board DART and the LEIA and LUKE cameras on board the CubeSat LICIACube (ASI: Agenzia Spaziale Italiana) have been published \citep{Daly2023}. 
\cite{Daly2023} report a set of freshly estimated values for the critical parameters in this study: the mass of the system, $(5.6\pm 0.5)\times 10^{11}$ kg; the equatorial extents, $850\pm 5$ m on average; the equivalent diameter of both Didymos, $761\pm 26$ m, and Dimorphos, $151\pm 5$ m (uncertainties are given at $1-\sigma$ level). Therefeore, the mass of Didymos can also be easily estimated, assuming equal density for the two bodies. We are conscious that the Didymos shape model is preliminary, however, such set of parameters tends to confirm the possibility of mass lift off from the equatorial region of Didymos.  In fact, even in the worst case compatible with available uncertainty ranges (largest system mass, smallest  Dimorphos equivalent diameter, largest Didymos equivalent diameter, and smallest Didymos equatorial extent), the ratio between the centripetal and gravitational acceleration at the equator, $a_c/g=\omega^2r_{eq}^3/(GM_P)$ is larger than 1 (1.07), indicating outwards acceleration.  $a_c/g$ is the driving magnitude here (rather than density), where $\omega$ is the spin rate and $r_{eq}$ is the equatorial radius of Didymos. The nominal values from \cite{Daly2023} give  $a_c/g=1.235$, which is very close to the $(M_1, V_4)$ set, providing more mass in orbit than our nominal case. 
$a_c/g=1$ for $r_{eq}^{\prime}=396$ m, which is the minimum distance from the spin axis at which lift off is possible, some $30$ m smaller than the estimated equatorial radius extent.

It is interesting to notice that the estimated bulk density is reported as $2400\pm 300$ kg/m$^3$, a value that would inhibit mass shedding on a sphere spinning at the Didymos spin rate. 
The bulk density of a body is the critical parameter only in the case in which its shape is fixed. Comparing a sphere with an equal mass oblate spheroid with equatorial axis $a$ and shortest axis $c$, the corresponding critical spin rate is 

\begin{equation}
\omega_{cr}^{\prime}=\sqrt{\frac{4 \pi G \rho^{\prime}}{3}\frac{c}{a}}
\end{equation}

\noindent
At equal critical spin rate, the relationship between the density of the oblate shaped and the spherical body is $\rho^{\prime}/\rho=a/c$. 
E.g., for an oblate body with $a/c=1.37$ (the case of Didymos, as its extent along spin axis is estimated to be $620\pm 15$ m by \cite{Daly2023}), the critical density is $\rho^{\prime}=2924$ kg/m$^3$, while, for an equal mass spherical Didymos, it would be $\rho=2134$ kg/m$^3$.

Indications of the presence of dust around Didymos can be envisioned by  looking at the LICIACube images. However, we prefer to avoid speculation at this point, and we delay further interpretation to a forthcoming companion paper. Once further estimate improvement of the main physical parameters and an updated Didymos shape model --pre-Hera mission-- will be available, a refined model will be carried out on the ``real'' Didymos system.

We also plan to study the interplay between mass ejected from Dimorphos by the DART impact and the stationary mass distribution around the system --if any. In fact, that may potentially affect the actual mass density  to be estimated by the Hera rendezvous with the Didymos system in 2027 and the calibration of instrumentation on board.

\section*{Acknowledgements}
NT, ACB and PBL acknowledge funding by the NEO-MAPP project through grant agreement 870377, in the frame of the EC H2020-SPACE-2019. NT acknowledges funding by CONICET. ACB and PBL acknowledge funding by the Ministerio de Ciencia Innovación (PGC 2018) RTI2018-099464-B-I00. FM acknowledges financial supports from grants MCIN/AEI/10.13039/ 501100011033 (GRANT PID2021-123370OB-I00),
European Union NextGeneratioEU/PTR, P18-RT-1854 from Junta de Andalucia, and CEX2021-001131-S funded by MCIN/AEI/10.13039/501100011033.

\end{document}